\documentclass[a4paper,11pt]{article}
\usepackage{pos}

\newcommand\figwidth{0.75}

\title{Localization at the quenched SU(3) phase transition}

\author*[a,b]{Tamas G.\ Kovacs}

\affiliation[a]{Department of Physics, Eotvos Lorand University,\\
  Pazmany Peter setany 1/a, 1117 Budapest, Hungary}

\affiliation[b]{Institute for Nuclear Research,\\
Bem ter 18/c, 4026 Debrecen, Hungary}

\emailAdd{tamas.gyorgy.kovacs@ttk.elte.hu}

\abstract{It is known that the deconfining transition of QCD is accompanied by
  the appearance of localized eigenmodes at the low end of the Dirac
  spectrum. In the quenched case localization appears exactly at the critical
  temperature of deconfinement. In the present work, using quenched
  simulations exactly at the critical temperature, we show that the
  localization properties of low Dirac modes change abruptly between the
  confined and deconfined phase. This means that in the real Polyakov loop
  sector, the mobility edge has a discontinuity at the critical
  temperature. In contrast, in the complex sector, there is no such
  discontinuity at $T_c$, even the lowest Dirac modes remain delocalized at the
  critical temperature in the deconfined phase. }

\FullConference{%
 The 38th International Symposium on Lattice Field Theory, LATTICE2021
  26th-30th July, 2021
  Zoom/Gather@Massachusetts Institute of Technology
}

\begin{document}
\maketitle

\section{Introduction}

The spectrum of the QCD Dirac operator is well known to encode many important
features of the underlying gauge theory. In particular, it undergoes dramatic
changes at the finite temperature transition into the quark-gluon plasma
state. In the present work we study how certain properties of the Dirac
spectrum change at the finite temperature phase transition of the pure $SU(3)$
gauge theory. The pure gauge theory provides a good testing ground for this
study, since -- in contrast to QCD with physical quark masses -- this system
has a genuine first order phase transition, not just a crossover. As the
character of the gluon fields changes dramatically at the phase transition,
the corresponding changes in the Dirac spectrum can be easily detected.

The phase transition of the quenched theory is signaled by the spontaneous
breaking of the $Z(3)$ symmetry of the Polyakov loop. While in the low
temperature, confining (symmetric) phase the expectation of the Polyakov loop
is zero, above the phase transition temperature, in the deconfined phase, the
symmetry is spontaneously broken and the Polyakov loop develops a nonzero
expectation value, belonging to one of the three $Z(3)$ sectors. In the present
work we perform quenched simulations exactly at the phase transition
temperature, where in a finite volume the system can tunnel between the
confined and deconfined phase. Therefore, in a simulation performed at the
transition temperature we can sample both phases, and in fact, in the
deconfined phase all three Polyakov loop sectors as well.

\begin{figure}
\begin{center}
\includegraphics[width=\figwidth\columnwidth,keepaspectratio]{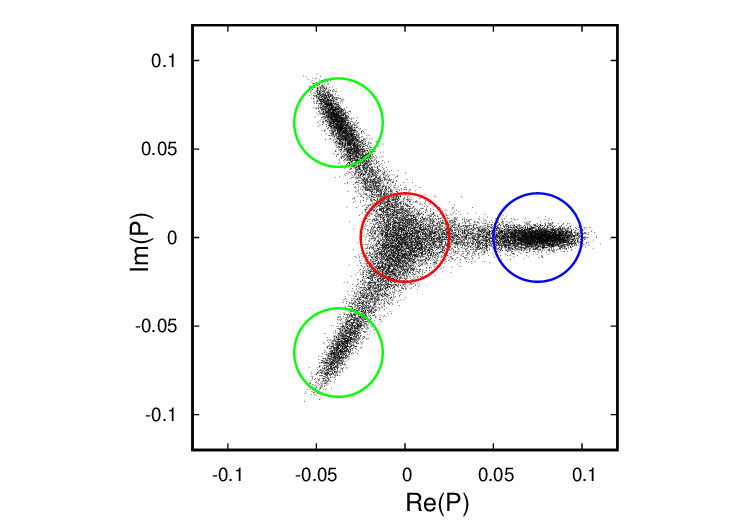}
\caption{\label{fig:plp488_b6.062} Scatter plot of the Polyakov loop in the
  complex plane. The encircled regions are the different sectors, the confined
  in the center (red circle), the deconfined real sector on the right (blue
  circle) and the deconfined complex sectors to the left (green circles). }
\end{center}
\end{figure}

Concerning the Dirac spectrum we have to distinguish three different sectors
in terms of the Polyakov loop, and the Dirac spectrum is expected to behave
differently in these sectors. These are the confined phase, the real sector in
the deconfined phase and the two complex sectors in the deconfined phase. Due
to the complex conjugation symmetry, Dirac spectra in the two complex sectors
are the same and we do not have to treat these sectors separately
\cite{Cardinali:2021fpu}. In Fig.~\ref{fig:plp488_b6.062} we depict the
different sectors by showing a scatter plot of the complex Polyakov loop on an
ensemble of $48^3\times 8$ lattice configurations exactly at the critical
temperature. We can see that although the probability of finding the Polyakov
loop in the four above mentioned sectors is enhanced, there is still
considerable tunneling among the different sectors.

\section{Details of the simulation}

Let us first summarize the technical details of the simulations. We generated
quenched gauge field configurations of spatial linear size $L=48$ and temporal
size $N_t=8$ using the Wilson gauge action at the critical value of the
coupling $\beta=6.063$. The configurations were separated into three different
groups (sectors) according to the value of the average Polyakov
loop. Configurations with $|P|<0.03$ were assigned to the confined phase, the
ones with $|P|>0.06$ were considered to be in the deconfined phase. The latter
configurations were further sorted into the real and complex Polyakov loop
sector according to the sign of the real part of the Polyakov loop . This set
of criteria was found to include the peaks of the Polyakov loop distribution
into the respective sectors (see Fig.~\ref{fig:plp488_b6.062}). We computed
the lowest 100 eigenvalues (with positive imaginary parts) of the overlap
Dirac operator separately on the three ensembles of configurations belonging
to the above three sectors.

\section{Results}

\begin{figure}
\begin{center}
\includegraphics[width=\figwidth\columnwidth,keepaspectratio]{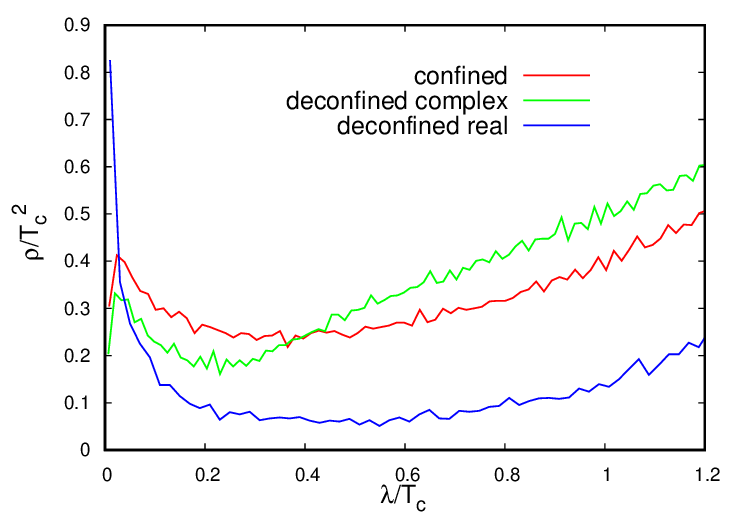}
\caption{\label{fig:spd488_b6.063} The spectral density of the overlap Dirac
  operator on the three ensembles of $48^3\times 8$ lattice configurations at
  Wilson $\beta=6.063$. }
\end{center}
\end{figure}

The simplest quantity characterizing the Dirac spectrum is the spectral
density. In Fig.~\ref{fig:spd488_b6.063} we compare the spectral density in
the three different sectors. The confined phase and the deconfined complex
sectors exhibit qualitatively similar spectral densities. In contrast, the
spectral density in the deconfined real sector is markedly different. It has a
high, but narrow spike at the very low end, above which the spectral density
drops considerably. The narrow spike in the spectral density can be attributed
to mixing instanton-antiinstanton zero modes that produce several pairs of
complex eigenvalues of rather small magnitude \cite{Vig:2021oyt}. Although in
the other two sectors there is also a slight accumulation of eigenvalues near
zero, this is by far less pronounced than in the real sector. This is caused
by a stronger mixing of topological zero modes, which in the confined phase is
due to a higher density of topological fluctuations. In the deconfined complex
Polyakov sectors this stronger mixing can be attributed to the more extended
nature of the topological zero modes
\cite{Kraan:1998pm}-\cite{Gattringer:2002tg}.

The difference in the bulk spectral density between the real and complex
sector in the confined phase can be qualitatively understood by considering
the lowest Matsubara modes in the free theory with different boundary
conditions. Indeed, the effective boundary condition for the Dirac equation is
a combination of the phase $\pi$, coming from the antiperiodic boundary
condition for fermions, and the phase of the Polyakov loop. In the real sector
the Polyakov loop phase is zero, in the complex sectors, it is $\pm 2\pi/3$,
making the magnitude of the overall effective phase $\pi$ and $\pi/3$,
respectively. Thus the lowest free Matsubara mode is much higher in the real
sector, and this makes the density of low-lying bulk modes considerably lower
also in the interacting case. In fact, this is how the breaking of the $Z(3)$
symmetry by dynamical quarks can be understood, since the quark determinant
favors the real sector that has fewer small eigenvalues.

\begin{figure}
\begin{center}
\includegraphics[width=\figwidth\columnwidth,keepaspectratio]{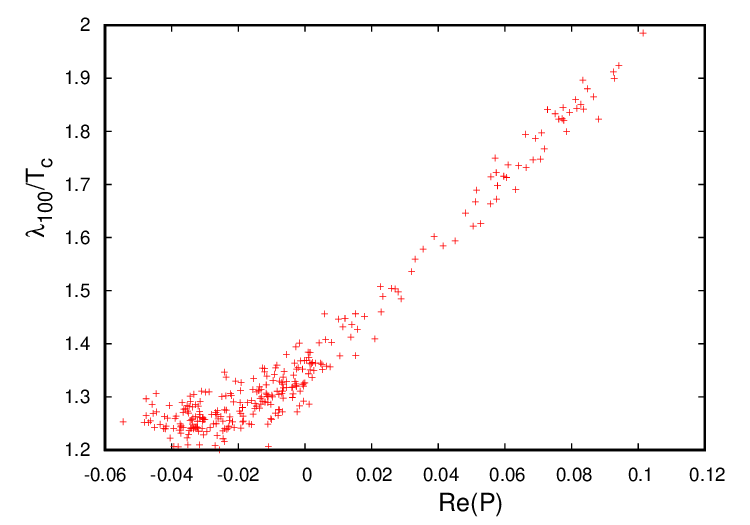}
\caption{\label{fig:plp_vs_ev} Scatter plot of the real part of the average
  Polyakov loop versus the $100^{th}$ smallest Dirac eigenvalue on a set of
  lattice configurations. }
\end{center}
\end{figure}

The direct connection between the Polyakov loop and the Dirac spectrum is
further demonstrated in Fig.~\ref{fig:plp_vs_ev}, where we plot the real
part of the average Polyakov loop versus the $100^{th}$ eigenvalue of the Dirac
operator. For positive real part there is a strong correlation between the two
quantities. It is clear that the more ordered the Polyakov loop is in the real
sector, the fewer low eigenvalues the Dirac operator has.

\section{Localization at $T_c$}

It is known that in the high temperature phase the lowest part of the Dirac
spectrum consists of localized eigenmodes (for a recent review see
\cite{Giordano:2021qav}). Approaching the transition temperature from above,
the mobility edge, separating the low localized modes from the bulk of the
spectrum, moves toward zero, and when it reaches zero, all modes become
delocalized, as expected in the low temperature phase. In the quenched case,
by extrapolating the mobility edge, it was found to vanish exactly at the
critical temperature \cite{Kovacs:2017uiz,Vig:2020pgq}. A naturally arising
question is whether the lowest part of the spectrum is localized or
delocalized exactly at the critical point. More precisely, the extrapolations
presented in Refs.~\cite{Kovacs:2017uiz,Vig:2020pgq} are compatible both with
a continuously vanishing mobility edge and one that has a discontinuity at the
phase transition. Our present simulation, done exactly at the critical point,
gives us an opportunity to distinguish these two possibilities.  This can be
done by looking at the localization properties of the lowest Dirac modes
separately in the confining and the deconfining phase and comparing the values
of the mobility edge in the two phases.

\begin{figure}
\begin{center}
\includegraphics[width=\figwidth\columnwidth,keepaspectratio]{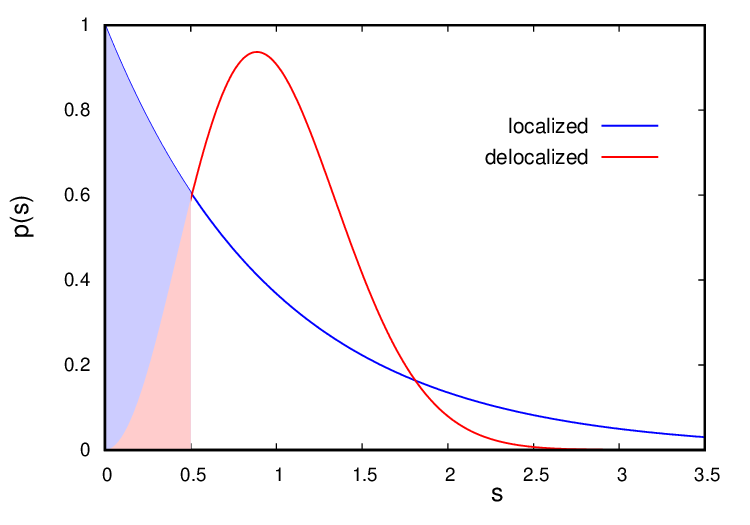}
\caption{\label{fig:lsd_locdeloc} The unfolded level spacing distribution,
  corresponding to localized and delocalized modes. The red shaded region
  depicts the integral, defining $I_{0.5}$ for the delocalized case, the blue
  one the difference of the integrals for the localized and the delocalized
  case.}
\end{center}
\end{figure}

The simplest way to decide whether eigenmodes in a certain region of the
spectrum are localized or delocalized is to compute the unfolded level spacing
distribution (ULSD). If the modes are localized, the corresponding eigenvalues
are statistically independent, they are described by a Poisson distribution,
and the unfolded level spacing distribution is exponential. If the eigenmodes
are delocalized, the ULSD is well approximated by the Wigner surmise of the
corresponding random matrix universality class, which in the case of the
overlap Dirac operator, is the unitary class. In Fig.~\ref{fig:lsd_locdeloc}
these two distributions are shown. A convenient way to monitor how the
distribution changes across the spectrum is to consider instead of the whole
distribution, just a single parameter that can distinguish the two extreme
cases (localized and delocalized). For this purpose we use the parameter
\begin{equation}
  I_{0.5} = \int_0^{0.5} p(x) d\!x,
\end{equation}
where the upper limit of the integration was chosen to be the crossing point
of the two distributions (approximately 0.5) to maximize the difference
between the two limiting cases (see Fig.~\ref{fig:lsd_locdeloc}).

\begin{figure}
\begin{center}
\includegraphics[width=\figwidth\columnwidth,keepaspectratio]{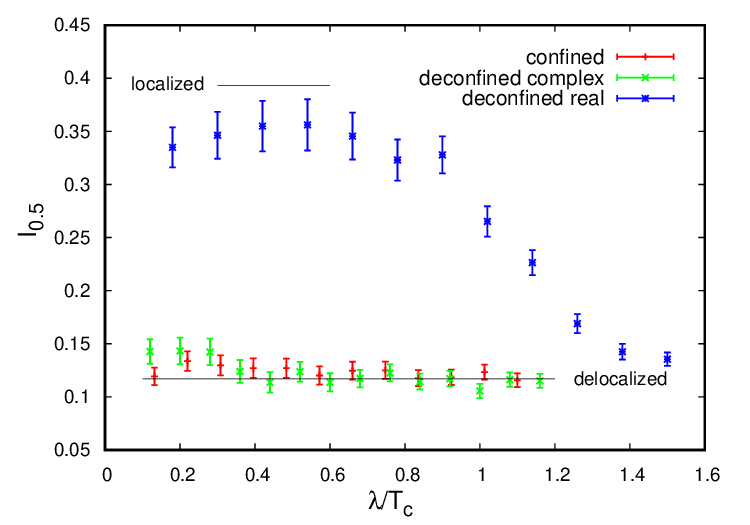}
\caption{\label{fig:uf488} The parameter $I_{0.5}$ across the spectrum
  computed separately in the three different sectors, the confined phase and in the
  deconfined phase in the real and the complex sector on the $48^3 \times 8$
  lattices. }
\end{center}
\end{figure}

To monitor how the ULSD changes across the spectrum and to determine the
mobility edge, we split the spectrum into narrow bins and in each bin
separately compute the parameter $I_{0.5}$. For details of the unfolding and
how we assign the eigenvalue pairs to bins, see the Appendix of
Ref.~\cite{Vig:2020pgq}. We repeated this procedure for all three sectors,
i.e.\ the confined phase and for the deconfined phase in the real and complex
sector. The results are shown in Fig.~\ref{fig:uf488}. It is clear from the
figure that both in the confined phase and in the deconfined complex sector,
already the lowest Dirac modes are delocalized, thus the mobility edge is
effectively at zero. However, in the deconfined real sector, in the one that
would be selected if dynamical fermions were present, there is a finite range
in the spectrum above zero where the eigenmodes seem to be localized. In this
region $I_{0.5}$ almost reaches the value expected for localized modes. The
slight deviation might be due to the finite system size or some more subtle
effects (see
Refs.~\cite{Alexandru:2019gdm}-\cite{Alexandru:2021xoi}). Notwithstanding
these details, it is obvious from the figure that the localization properties
of the lowest Dirac eigenmodes change abruptly between the confined phase and
the deconfined real Polyakov sector (the physical sector). This is consistent
with the fact that the transition in the $SU(3)$ pure gauge theory is first
order. It is, however, surprising that the deconfined complex Polyakov sector
does not exhibit such an abrupt change, the lowest modes there are
delocalized, like in the confined phase.

\section{Conclusions}

In the present paper we studied how the spectral properties of the Dirac
operator change at the finite temperature first order phase transition of the
pure $SU(3)$ gauge theory. We showed that the phase transition is accompanied
by discontinuous changes in the spectral properties. Namely, both the spectral
density around zero and the localization properties of the lowest modes change
abruptly at the transition. In particular, there is a band around zero in the
spectrum, where the corresponding eigenmodes become localized at the
transition in the real Polyakov loop sector. This also implies that the
mobility edge is discontinuous at the transition if we restrict our attention
to the real (physical) Polyakov loop sector.  In contrast, in the complex
sector even the lowest eigenmodes remain delocalized at the transition. This
property can be qualitatively understood by the more extended nature in this
sector of zero modes carried by Kraan-van Baal calorons.

\end{document}